\documentclass[superscriptaddress,pra]{revtex4}

\usepackage[latin1]{inputenc}
\usepackage{graphicx}
\usepackage{amsmath}
\usepackage{amssymb}
\usepackage{dsfont}
\usepackage{graphicx}

\makeatletter
\def\erf{\mathop{\operator@font erf}\nolimits}
\makeatother

\newcommand\be{\begin{equation}}
\newcommand\ee{\end{equation}}

\begin{document}

\title{NOON states in trapped ions}
\author{D. Rodr\'{\i}guez-M\'endez and H.M. Moya-Cessa} \affiliation{INAOE, Coordinaci\'on de Optica, Apdo. Postal 51 y
216, 72000 Puebla, Pue., Mexico}

\begin{abstract}
We show how NOON states may be generated in ion traps. We use the
 individual interaction of light with each of two vibrational
 modes of the ion to entangle them. This allows us to generate
 NOON states with  $N=8$.
\end{abstract}
\pacs{} \maketitle


\section{Introduction}
Nonclassical states  for mono-mode fields have attracted a great
deal of attention over the years because of their fundamental and
technological  value, among them we have 1) macroscopic quantum
superpositions of quasiclassical coherent states with different
mean phases or amplitudes \cite{1,2,2-5}, 2) squeezed states
\cite{3}, and 3) the particularly important limit of extreme
squeezing, i.e. Fock or number states \cite{4}. More recently some
bimodal fields have also attracted attention because they may be
used for high precision phase measurements: NOON states, i.e.,
nonclassical states of combined photon pairs  \cite{5,6}.  Many
applications in quantum imaging, quantum information and quantum
metrology \cite{7} depend on the availability of entangled photon
pairs because entanglement is a distinctive feature of quantum
mechanics that lies at the core of many new applications. These
maximally path-entangled multiphoton states may be written in the
form

\begin{equation}
\left| {N00N} \right\rangle _{a,b}  = \frac{1}{{\sqrt 2 }}\left(
{\left| N \right\rangle _a \left| 0 \right\rangle _b  + \left| 0
\right\rangle _a \left| N \right\rangle _b } \right).
\end{equation}

It has been pointed out that NOON states manifest unique coherence
properties by showing that they exhibit a periodic transition
between spatially bunched and antibunched states when   undergo
Bloch oscillations. The period of the bunching/antibunching
oscillation is N times faster than the period of the oscillation
of the photon density \cite{12}.

The greatest $N$ for which NOON states have been produced is $N=5$
\cite{5}. Most  schemes to generate  this class of states are
either for optical \cite{5,6} or microwave fields. In this
contribution, we would like to analyze the possibility to generate
them in ions \cite{wine,wine2,wine3,ion,Wall,Buzek,moya}, i.e.
NOON states of their vibrational motion. We will show that they
may be generated with $N=8$.

\section{Ion-laser interaction}
The Hamiltonian for an ion  in a two-dimensional Paul trap has the
form
\begin{equation}
\hat{H}=\nu_x a_x^{\dagger}a_x+\nu_y a_y^{\dagger}a_y
+\frac{\omega_{ee} }{2}\hat{A}_{ee}+
 \left( \hat{A}_{ge}\lambda E^{(-)}(\hat{x},\hat{y},t)+H.c.\right), \label{ion1}
\end{equation}
with $\lambda$ the electronic coupling matrix element,
$E^{(-)}(\hat{x},\hat{y},t)$ the negative part of the classical
electric field of the driving field. The operators $\hat{A}_{jk}$
take into account the transitions between the states $|j\rangle$
and $|k\rangle$ ($g$ for ground and $e$ for excited). We assume
the ion driven by a plane wave
\begin{equation}
E^{(-)}(\hat{x},\hat{y},t)=E_0e^{-i(k_x \hat{x}+k_y \hat{y}+\omega
t)}
\end{equation}
with $k_j, j=x,y$ the wavevectors of the driving field and define
the Lamb-Dicke parameters
\begin{equation}
\eta_x=2\pi\frac{\sqrt{_x\langle
0|\Delta\hat{x}^2|0\rangle}_x}{\lambda_x}, \qquad
\eta_y=2\pi\frac{\sqrt{_y\langle
0|\Delta\hat{y}^2|0\rangle_y}}{\lambda_y},
\end{equation}
such that we redefine
\begin{equation}
k_x \hat{x}=\eta_x(a_x+a_x^{\dagger}), \qquad k_y
\hat{y}=\eta_y(a_y+a_y^{\dagger}).
\end{equation}
In the resolved sideband limit, the vibrational frequencies
$\nu_x$ and $\nu_y$ are much larger than other characteristic
frequencies and the interaction of the ion with the two lasers can
be treated separately, using a nonlinear Hamiltonian
\cite{matos1,matos2}. The Hamiltonian (\ref{ion1}) in the
interaction picture can then be written as
\begin{equation}\label{}
    H_I=\Bigg\{ \begin{array}{c}
        \Omega_x^{(k)}e^{-\eta_x^2/2}\hat{A}_{eg}\frac{\hat{n}_x!}{(\hat{n}_x+k)!}L_{\hat{n}_x}^{(k)}(\eta_x^2)a_x^k +H.c, \qquad \eta_y=0,\qquad \delta=k\nu_x \\
         \\
        \Omega_y^{(k)}e^{-\eta_y^2/2}\hat{A}_{eg}\frac{\hat{n}_y!}{(\hat{n}_y+k)!}L_{\hat{n}_y}^{(k)}(\eta_y^2)a_y^k +H.c, \qquad  \eta_x=0,\qquad \delta=k\nu_y \\
      \end{array}
\end{equation}
where $L_{\hat{n}_j}^{(k)}, j=x,y$ are the operator-valued
associated Laguerre polynomials, the $\Omega$'s are the Rabi
frequencies and $\hat{n}_j=a_j^{\dagger}a_j, j=x,y$. If we
consider consider $\eta_y=0$ and $\eta_x\ll 1$ and take
$\delta=4\nu_x$ we obtain
\begin{equation}\label{}
    H_I^{(x)}\approx gA_{21}a_x^4+H.c.
\end{equation}
with $g=\Omega_x\frac{\eta_x^4}{4!}$. The evolution operator for
the interaction Hamiltonian is then (in the atomic basis, see
\cite{Phoenix})
\begin{equation}\label{}
    U_I^{(x)}(t)=e^{-iH_I^{(x)}t}=
     \left( {\begin{array}{*{20}c}
   C_{\hat{n}}^x &  - iS_{\hat{n}}^{\dagger x}V_x^4  \\
   -iV_x^{ \dagger 4}S_{\hat{n}}^x &   C_{\hat{n}-4}^x
\end{array}} \right), \label{evola}
\end{equation}
where
\begin{equation}
\begin{array}{c}
\nonumber
C_{\hat{n}}^x=\cos(\sqrt{(\hat{n}_x+4)(\hat{n}_x+3)(\hat{n}_x+2)(\hat{n}_x+1)}gt),
\\
S_{\hat{n}}^x
=\sin(\sqrt{(\hat{n}_x+4)(\hat{n}_x+3)(\hat{n}_x+2)(\hat{n}_x+1)}gt),
\end{array}
\end{equation}
and the operator
\begin{equation}
V_x=\frac{1}{\sqrt{\hat{n}_x+1}}a_x
\end{equation}
the Susskind-Glogower (phase) operator \cite{Suskind}.
\section{Generation of NOON states}
By  starting with the ion in the excited state and the initial
vibrational state in the vacuum state, i.e.
$|0\rangle_x|0\rangle_y$, if we set $\eta_y=0$, after a convenient
time, this is, the time when the probability to find the ion in
its excited state is zero (meaning that the ion, by passing from
its excited to its ground state, gives $4$ phonons to the
vibrational motion),  we can generate the state
$|4\rangle_x|0\rangle_y$. Repeating this procedure (with the ion
reset again to the excited state, via a rotation), but now with
$\eta_x=0$, for phonon are added to the $y$-vibrational motion,
generating the two-dimensional state $|4\rangle_x|4\rangle_y$.

 Therefore, if we consider the ion initially in a superposition of ground and
excited states, and the $|4\rangle_x|4\rangle_y$ vibrational state
\begin{equation}
|\psi_{init}\rangle=\frac{1}{\sqrt{2}}(|e\rangle+|g\rangle)|4\rangle_x|4\rangle_y,
\end{equation}
for $\eta_y=0$ and $\tau_p$, the state generated is
\begin{equation}
|\psi_{\eta_y=0}\rangle=\frac{i}{\sqrt{2}}(|e\rangle|0\rangle_x+|g\rangle|8\rangle_x)|4\rangle_y.
\end{equation}
Now, we consider this state as initial state for the next
interaction with $\eta_x=0$ and still the interaction time
$\tau_p$, to produce
\begin{equation}
|\psi_{\eta_x=0}\rangle=-\frac{1}{\sqrt{2}}(|e\rangle|0\rangle_x|8\rangle_y+|g\rangle|8\rangle_x|0\rangle_y).
\end{equation}
Next, the ion is rotated via a classical field (an on-resonance
interaction) such that the state

\begin{equation}
|\psi_R\rangle=-\frac{1}{{2}}\left[|e\rangle(|0\rangle_x|8\rangle_y-|8\rangle_x|0\rangle_y)
+|g\rangle(|0\rangle_x|8\rangle_y+|8\rangle_x|0\rangle_y\right]
\end{equation}
is obtained. Finally by measuring the ion in its excited state we
produce the NOON state
\begin{equation}
|NOON\rangle_e=\frac{1}{\sqrt{2}}(|0\rangle_x|8\rangle_y-|8\rangle_x|0\rangle_y),
\end{equation}
and if the ion is measured in the ground state, also a NOON state
is produced:
\begin{equation}
|NOON\rangle_g=\frac{1}{\sqrt{2}}(|0\rangle_x|8\rangle_y+|8\rangle_x|0\rangle_y),
\end{equation}
\section{Conclusions}
It has been shown a way to produce high NOON states ($N=8$) by
entangling the vibrational motion of an ion trapped in two
dimensions. The entanglement is produced by a set of interactions
of the trapped ion with  laser fields conveniently tuned to
produce  $4$-phonons transitions. It should be noted that, because
the Lamb-Dicke regime is assumed, the four-phonon transitions we
are proposing is difficult to achieve. Decoherence processes are
usually not taken into account in ion-laser interactions because
these interactions  are, in general, not affected by the
environment. However, due to the interaction times needed to
generate the states presented here, they  may be considered in
this case. Here however we have only treated the ideal case as a
first approach. We will study the consequences of the environment
elsewhere.

\end{document}